\documentclass[sigconf]{acmart}
\usepackage{amsthm}
\usepackage{amsmath}
\usepackage{amsmath}
\usepackage{algorithm}
\usepackage{algpseudocode}
\usepackage{enumitem}

\usepackage{array}
\usepackage{eqparbox}
\usepackage{url}

\makeatletter
\g@addto@macro{\UrlBreaks}{\UrlOrds}
\makeatother

\AtBeginDocument{%
  \providecommand\BibTeX{{%
    \normalfont B\kern-0.5em{\scshape i\kern-0.25em b}\kern-0.8em\TeX}}}
\copyrightyear{2024}
\acmYear{24}

\begin{document}

\title{Tortoise: An Authenticated Encryption Scheme}

\author{Kenneth Odoh}
\affiliation{\institution{\url{https://kenluck2001.github.io}}}
\email{kenneth.odoh@gmail.com}

\begin{abstract}

Given the open nature of the Internet, there is a need for authentication schemes to address inherent trust issues. We present Tortoise, an experimental nonce-based authenticated encryption scheme modeled on the Synthetic Counter-in-Tweak. This paper demonstrates a generalizable plug-and-play framework for converting block cipher into Authenticated Encryption with Associated Data. As part of this work, we utilized an XOR procedure for constructing a generic tweakable cipher. Finally, we support two modes: nonce-respecting and nonce-misuse-resistant.\\\\ 
\textbf{Source code} available at \\\url{https://github.com/kenluck2001/cipherResearch/tree/main/src/tortoise}.

\end{abstract}



\maketitle

\section{Introduction}

The Internet provides an open platform for several forms of interaction between users. Given the open nature of the web, there are trust issues due to bad actors taking advantage of unsuspecting users. Cryptography-based systems can provide the right tool for achieving trust. There are two broad categories of cryptosystems: Symmetric-Key encryption and Asymmetric-Key encryption~\cite{Diffie2006}. The encryption and decryption of messages in Symmetric-Key encryption use the same key and a pair of keys (private and public key) for encryption and decryption in Asymmetric-Key encryption.

Authenticated encryption (AE) allows multiple parties to exchange messages with secrecy and integrity. Users can verify the real identity of the message creator and prevent message forgeries that can erode trust. The present-day electronic transactions are only possible partly due to the invention of secure authenticated encryption schemes. Our work uses the Symmetric-Key encryption primitive for building our authenticated encryption scheme, which is different for digital signatures~\cite{Diffie2006} that focuses on Asymmetric-Key encryption.

Recent CAESER competitions~\cite{caesar} have shown the need for authenticated encryption schemes. Our work introduced an Authenticated Encryption with Associated Data (AEAD) scheme based on well-known cipher primitives with well-studied security characteristics. Tortoise is a novel cryptographic algorithm based on a tweakable variant of AES~\cite{aes2001}. Our proposed scheme provides a generalizable framework capable of converting any block cipher (slightly modified to accommodate different block sizes) into an authenticated encryption scheme.

Tortoise is a symmetric cryptographic scheme that allows parties to communicate by encrypting and authenticating their messages. AEAD provides confidentiality, integrity, and authentication. Tortoise uses 128 bits for message blocks. However, this tweakable cipher block size is pre-determined by the message block size of the underlying cipher, which in our case is AES~\cite{aes2001}. The algorithm is easy to implement in both hardware and software. The nonce-resistant variant of Tortoise provides full misuse-resistant authenticated encryption (MRAE) security. Furthermore, our design choices follow the expectation of long-term cipher deployment on edge devices that are unmaintained after installation. In this situation, ciphers subject to rigorous security review over an extended period are advantageous for deployment.

This paper presents a real-life demonstration of our Tortoise cipher organized as follows. We did some literature review in Section~\ref{related-work}, an overview of our implementation in Section~\ref{system-overview}, and security analysis in Section~\ref{security-analysis}. Finally, we provide limitations, future work, and conclusions in Section~\ref{limitations-and-future-work} and Section~\ref{conclusions}.

\section{Contributions}
\label{contributions}

Our contributions are as follows:

\begin{itemize}[topsep=7pt, partopsep=0pt, listparindent=0pt, leftmargin=5pt]
\item We have created a generalizable framework for converting any block cipher into an authenticated encryption cipher. Two modes are supported: nonce-respecting and nonce-misuse-resistant.

\item Our scheme shows a framework for tweaking any block cipher.
\end{itemize}

\footnote{This work arose from independent research done in this blog post: \url{https://kenluck2001.github.io/blog_post/probing_real-world_cryptosystems.html}}

\section{Related Work}
\label{related-work}

Synthetic Counter-in-Tweak (SCT)~\cite{Peyrin2016} is a combination of Encrypted Parallel Wagman-Carter (EPWC) and Counter-in-Tweaks (CTRT). As a result, it provides security for nonce-respecting mode beyond birthday bounds. On the other hand, in the nonce-misuse resistance, security is provided until the birthday-bound period~\cite{Fleischmann2012}.

The Tweakey framework~\cite{Jean2014} is used in Deoxys~\cite{Jean2016} and limits the flexibility of the encryption round. This setup is undesirable because of tight coupling with the underlying encryption routine. On the contrary, the XOR scheme does not modify the encryption round. Hence, we can use it as a foundation to create a generic framework for tweaking any block cipher.

EAX ~\cite{Bellare2003} is an authenticated scheme that introduced the use of associated data with provable security in AE. This is further used in Deoxys~\cite{Jean2016}, Ascon ~\cite{Dobraunig2016}. Nonce reuse schemes are frequent in many authentication schemes. However, HBS and BTM provide nonce-misuse resistance ~\cite{Fleischmann2012} as supported in Deoxys~\cite{Jean2016}. Our authentication scheme follows authenticate-then-encrypt, which was proven secure using a tweakable cipher or CBC mode~\cite{Krawczyk2001}.

\section{System Overview}
\label{system-overview}
In this section, we have provided a rationale for our design in Subsection~\ref{design-rationale}, specification in Subsection~\ref{design-rationale}, a discussion of tweakable cipher in Subsection~\ref{tweakable-cipher}, and implementation in Subsection~\ref{implementation} respectively.

\subsection{Design Rationale}
\label{design-rationale}

We have taken a conservative approach to building our cipher based on principles with well-understood security properties. Our design entails constructing a novel authenticated encryption scheme based on the well-tested SCT ~\cite{Peyrin2016} primitive as a foundation, then employing a tweakable cipher based on the AES cipher. The resulting AEAD handles arbitrary message lengths using the padding scheme for multiples of message block size, $n$, using PKCV7 strategy~\cite{Kaliski1998}. Our work follows Definition 1 and accepts associated data as input to encryption and decryption flow. ASCON~\cite{Dobraunig2016} and Deoxys~\cite{Jean2016} also use the associated data pattern in their construction.

\textbf{Definition 1}~\cite{Liskov2002}. Let $C, tag = E_A(P, K, A, N) $ be an authenticated encryption function, and $P = D_A(C, K, A, tag, N) $ be an authenticated decryption function where $P$: plaintext, $C$: ciphertext, $K$: key, $A$: associated data, and $N$: nonce respectively. Decryption works once when tags are valid and match, implying no forgery.

\subsection{Specification}
\label{specification}

Our work shares similarities to Deoxys~\cite{Jean2016} for using the same SCT ~\cite{Peyrin2016} formulation, but it is different too in peculiar ways. Deoxys~\cite{Jean2016} follows the Tweakey framework for tweaking block cipher with a dependency on a key expansion scheme suited for AES-like only cipher, thereby affecting generalizability to certain ciphers like Quantum-safe ciphers. On the contrary, we have opted for the XOR procedure with a 2-universal hashing function ~\cite{Liskov2002}.

Reusing nonce is a common issue in real-world cryptosystems. It is unavoidable in some cases as it may be impossible to keep track of used values as system restarts may reset the system. Once the nonce gets reused in a nonce-respecting mode, then a security vulnerability results.

\subsection{Tweakable Cipher}
\label{tweakable-cipher}

The tweak is desirable when changing the key is more expensive than changing tweaks. The tweak in our formulation has similar characteristics to CBC mode. A tweakable block cipher is secure even if the adversary knows the chosen tweak. It is conceptually similar to how initialization vectors get utilized in CBC mode.

\begin{itemize}[topsep=7pt, partopsep=0pt, listparindent=0pt, leftmargin=5pt]
\item The underlying base cipher is as follows: $c = E(p, k)$, $p = D(c, k)$ where $c$: ciphertext, $p$: plaintext, $k$: key.
\item The supported operations of the Tweakable cipher as shown: $C = E_n(K, T, P)$ is an encryption function that accepts key, $K$, tweak, $T$, and plaintext, $P$, then outputs ciphertext, $C$. $P = D_n(K, T, C)$ is an encryption function that accepts key, $K$, tweak, $T$, and ciphertext, $C$, then outputs plaintext, $P$.
\end{itemize}
Our construction of a tweakable cipher for Tortoise follows Theorem 1 with the proof shown on paper ~\cite{Liskov2002, Shoup1996}.

\textbf{Theorem 1}~\cite{Liskov2002}. Let $E_n(K, T, P) = E(P, h(T)) \oplus h(T) $ where $h(T)$ is chosen from the family of $\epsilon-\emph{AXU}_{2}$.
We used SHAKE128~\cite{SHA2015} as our hashing function, $h$.

\subsection{Implementation}
\label{implementation}

In our formulation, the tweak block size must match the message block size, which is a deviation from Deoxys~\cite{Jean2016}. We retain the 4-bit prefix used in Deoxys~\cite{Jean2016} and the same modes: nonce-respecting and nonce-misuse resistant.

The nonce-respecting procedures are available in Algorithm~\ref{encalgo-nonce-respecting} and Algorithm~\ref{decalgo-nonce-respecting}.

\begin{algorithm}[H]
\caption{Encryption Algorithm, $E_A(P, K, A, N)$}
\label{encalgo-nonce-respecting}
\begin{algorithmic}

\Require $P$: plainText, $K$: key, $A$: associated data, and $N$: nonce where $l_a$: length of associated data, $A$, of multiple of blocks of size, $n$ i.e $|A_i| = n$, $l_p$: length of plainText, $P$, of multiple of blocks of size, $n$ i.e $|P_i| = n$

\Ensure $C$: cipherText i.e $|C_i| = n$, $tag$

\State $auth = 0$
\State \Comment{// Processing associated data}
\ForAll{$i \in l_a$ }
    \State $auth = auth \oplus E_n(K, 0010|i, A_i)$ 
\EndFor

\State \Comment{// Processing plaintext data}
\State $checksum = 0^n$
\ForAll{$j \in l_p$ }
    \State $checksum = checksum \oplus P_j$ 
    \State $C_j = E_n(K, 0000|N|j, P_j)$ 
\EndFor

\State $fTag = E_n(K, 0001|N|l_p, checksum)$ 
\State \Comment{// Tag generation}

\State $tag = fTag \oplus auth$ 

\end{algorithmic}
\end{algorithm}

\begin{algorithm}[H]
\caption{Decryption Algorithm, $D_A(C, K, A, tag, N)$}
\label{decalgo-nonce-respecting}
\begin{algorithmic}

\Require $C$: cipherText, $K$: key, $A$: associated data, tag, and $N$: nonce where $l_a$: length of associated data, $A$, of multiple of blocks of size, $n$ i.e $|A_i| = n$, $l_c$: length of cipherText, $P$, of multiple of blocks of size, $n$ i.e $|P_i| = n$

\Ensure $P$: plainText i.e $|P_i| = n$, $\hat{tag}$

\State $auth = 0$
\State \Comment{// Processing associated data}
\ForAll{$i \in l_a$ }
    \State $auth = auth \oplus E_n(K, 0010|i, A_i)$ 
\EndFor

\State \Comment{// Processing ciphertext data}
\State $checksum = 0^n$
\ForAll{$j \in l_c$ }
    \State $P_j = D_n(K, 0000|N|j, C_j)$ 
    \State $checksum = checksum \oplus P_j$ 
\EndFor

\State $fTag = E_n(K, 0001|N|l_c, checksum)$ 
\State \Comment{// Tag generation}
\State $\hat{tag} = fTag \oplus auth$ 

\If{$\hat{tag} == {tag}$} 
    \Return plaintext, $P$
\Else
    \State \Comment{// Do nothing}
\EndIf 

\end{algorithmic}
\end{algorithm}

The nonce-misuse-resistant procedure as shown in Algorithm~\ref{encalgo-nonce-misuse-resistant} and Algorithm~\ref{decalgo-nonce-misuse-resistant} respectively.

\begin{algorithm}[H]
\caption{Encryption Algorithm, $E_A(P, K, A, N)$ as shown in Definition 1}
\label{encalgo-nonce-misuse-resistant}
\begin{algorithmic}

\Require $P$: plainText, $K$: key, $A$: associated data, and $N$: nonce where $l_a$: length of associated data, $A$, of multiple of blocks of size, $n$ i.e $|A_i| = n$, $l_p$: length of plainText, $P$, of multiple of blocks of size, $n$ i.e $|P_i| = n$

\Ensure $C$: cipherText i.e $|C_i| = n$, $tag$

\State \Comment{// Processing associated data}
\State $auth = 0$
\ForAll{$i \in l_a$ }
    \State $auth = auth \oplus E_n(K, 0010|i, A_i)$ 
\EndFor

\State \Comment{// Processing plaintext data}
\State $tag = auth$
\ForAll{$j \in l_p$ }
    \State $tag = tag \oplus E_n(K, 0000|N|j, P_j)$ 
\EndFor

\State \Comment{// Tag generation}
\State $tag = E_n(K, 0001|0^{4}|N, tag)$ 

\State \Comment{// Message encryption}
\ForAll{$j \in l_p$ }
    \State $C_j = P_j \oplus E_n(K, tag \oplus j, 0^{8}|N)$ 
\EndFor

\end{algorithmic}
\end{algorithm}

\begin{algorithm}[H]
\caption{Decryption Algorithm, $D_A(C, K, A, tag, N)$}
\label{decalgo-nonce-misuse-resistant}
\begin{algorithmic}

\Require $C$: cipherText, $K$: key, $A$: associated data, tag, and $N$: nonce where $l_a$: length of associated data, $A$, of multiple of blocks of size, $n$ i.e $|A_i| = n$, $l_c$: length of cipherText, $P$, of multiple of blocks of size, $n$ i.e $|P_i| = n$

\Ensure $P$: plainText i.e $|P_i| = n$, $\hat{tag}$

\State \Comment{// Message encryption}
\ForAll{$j \in l_p$ }
    \State $P_j = C_j \oplus E_n(K, tag \oplus j, 0^{8}|N)$ 
\EndFor

\State \Comment{// Processing associated data}
\State $auth = 0$
\ForAll{$i \in l_a$ }
    \State $auth = auth \oplus E_n(K, 0010|i, A_i)$ 
\EndFor

\State \Comment{// Processing plaintext data}
\State $\hat{tag} = auth$
\ForAll{$j \in l_p$ }
    \State $\hat{tag} = \hat{tag} \oplus E_n(K, 0000|N|j, P_j)$ 
\EndFor

\State \Comment{// Tag generation}
\State $\hat{tag} = E_n(K, 0001|0^{4}|N, \hat{tag})$ 

\State \Comment{// Tag verification}
\If{$\hat{tag} == {tag}$} 
    \Return plaintext, $P$
\Else
    \State \Comment{// Do nothing}
\EndIf 

\end{algorithmic}
\end{algorithm}

\section{Security Analysis}

\label{security-analysis}

We have adopted a conservative approach to our cipher design by basing the foundation on well-tested and verified security primitives and benefiting from the composability of these primitives in our unique construction. As a result, we can have minimal doubt that we can demonstrate the security of our construction without a very elaborate security analysis.

SCT~\cite{Peyrin2016} has proven its security properties by serving as the foundation for ASCON~\cite{Dobraunig2016} and Deoxys~\cite{Jean2016}. The underlying cipher is AES (immaterial to our authentication routine). The cipher choice in this work is arbitrary, and we focus on building a plug-and-play framework for converting any block cipher into an authenticated encryption scheme, provided that the tweak size is equal to the message block size. The proof of security properties of our tweaking scheme used in our formulation in Theorem 2 of paper ~\cite{Liskov2002}. We have provided evidence of a strong tweakable block cipher.

The paper specifically used AES, but any block cipher would suffice. Hence, we can revisit the security of AES~\cite{aes2001} with a track record of subjection to rigorous real-world security analysis. We can claim security levels of $2^{128}$ for AES 128. Our underlying tweakable AES~\cite{aes2001} is reasonably resilient to differential and linear cryptanalysis. However, it may be vulnerable to the meet-in-the-middle attack ~\cite{Bar-On2018}. More investigations are needed to evaluate resilience to related-key attacks.

\section {Limitations and Future Work}

\label {limitations-and-future-work}

We have identified some areas for improvement and limitations as follows:

\begin{itemize}[topsep=7pt, partopsep=0pt, listparindent=0pt, leftmargin=5pt]

\item We hope to improve this work in the future by providing hardware implementation.

\item We can extend this framework to develop generic AEADs such as quantum-safe authenticated ciphers.

\item Future work by relaxing the notion of MRAE in the form of Online AE (OAE), which requires only a single pass for nonce-misuse resistant ciphers. This planned update can use ideas from Romulus-M~\cite{GuoNist}, and we can create an improved nonce-misuse resistant cipher.

\item Our implementation did not consider mitigation for timing or side-channel attacks.

\end{itemize}

\section{Conclusions}

\label{conclusions}

We have released Tortoise as a general-purpose scheme for converting any block into AEAD. As a result of this work, we have created a custom XOR procedure for constructing a generic tweakable cipher.

\bibliographystyle{ACM-Reference-Format}
\bibliography{sample-sigconf}


\begin{thebibliography}{16}


\ifx \showCODEN    \undefined \def \showCODEN     #1{\unskip}     \fi
\ifx \showDOI      \undefined \def \showDOI       #1{#1}\fi
\ifx \showISBNx    \undefined \def \showISBNx     #1{\unskip}     \fi
\ifx \showISBNxiii \undefined \def \showISBNxiii  #1{\unskip}     \fi
\ifx \showISSN     \undefined \def \showISSN      #1{\unskip}     \fi
\ifx \showLCCN     \undefined \def \showLCCN      #1{\unskip}     \fi
\ifx \shownote     \undefined \def \shownote      #1{#1}          \fi
\ifx \showarticletitle \undefined \def \showarticletitle #1{#1}   \fi
\ifx \showURL      \undefined \def \showURL       {\relax}        \fi
\providecommand\bibfield[2]{#2}
\providecommand\bibinfo[2]{#2}
\providecommand\natexlab[1]{#1}
\providecommand\showeprint[2][]{arXiv:#2}

\bibitem[\protect\citeauthoryear{??}{aes}{2001}]%
        {aes2001}
 \bibinfo{year}{2001}\natexlab{}.
\newblock \bibinfo{title}{{Advanced Encryption Standard (AES)}}.
\newblock
  \bibinfo{howpublished}{\url{https://nvlpubs.nist.gov/nistpubs/FIPS/NIST.FIPS.197-upd1.pdf}}.
\newblock
\newblock
\shownote{Date accessed: July 28, 2023.}


\bibitem[\protect\citeauthoryear{??}{cae}{2001}]%
        {caesar}
 \bibinfo{year}{2001}\natexlab{}.
\newblock \bibinfo{title}{{CAESAR: Competition for Authenticated Encryption:
  Security, Applicability, and Robustness}}.
\newblock
  \bibinfo{howpublished}{\url{https://competitions.cr.yp.to/caesar.html}}.
\newblock
\newblock
\shownote{Date accessed: September 26, 2023.}


\bibitem[\protect\citeauthoryear{??}{SHA}{2015}]%
        {SHA2015}
 \bibinfo{year}{2015}\natexlab{}.
\newblock \bibinfo{title}{{SHA-3 Standard: Permutation-Based Hash and
  Extendable-Output Functions}}.
\newblock
  \bibinfo{howpublished}{\url{https://nvlpubs.nist.gov/nistpubs/FIPS/NIST.FIPS.202.pdf}}.
\newblock
\newblock
\shownote{Date accessed: July 28, 2023.}


\bibitem[\protect\citeauthoryear{{Bar-On, Achiya and Dunkelman, Orr and Keller,
  Nathan and Ronen, Eyal and Shamir, Adi}}{{Bar-On, Achiya and Dunkelman, Orr
  and Keller, Nathan and Ronen, Eyal and Shamir, Adi}}{2018}]%
        {Bar-On2018}
\bibfield{author}{\bibinfo{person}{{Bar-On, Achiya and Dunkelman, Orr and
  Keller, Nathan and Ronen, Eyal and Shamir, Adi}}.}
  \bibinfo{year}{2018}\natexlab{}.
\newblock \showarticletitle{{Improved Key Recovery Attacks On Reduced-Round AES
  with Practical Data and Memory Complexities}}. In
  \bibinfo{booktitle}{\emph{{Proceedings of the 38th Annual International
  Cryptology Conference on Advances in Cryptology}}}.
  \bibinfo{pages}{185--212}.
\newblock


\bibitem[\protect\citeauthoryear{Bellare, Rogaway, and Wagner}{Bellare
  et~al\mbox{.}}{2003}]%
        {Bellare2003}
\bibfield{author}{\bibinfo{person}{Mihir Bellare}, \bibinfo{person}{Phillip
  Rogaway}, {and} \bibinfo{person}{David~A. Wagner}.}
  \bibinfo{year}{2003}\natexlab{}.
\newblock \showarticletitle{EAX: A Conventional Authenticated-Encryption Mode}.
\newblock \bibinfo{journal}{\emph{IACR Cryptology ePrint Archive}}
  \bibinfo{volume}{2003} (\bibinfo{year}{2003}), \bibinfo{pages}{69}.
\newblock


\bibitem[\protect\citeauthoryear{Diffie and Hellman}{Diffie and
  Hellman}{2006}]%
        {Diffie2006}
\bibfield{author}{\bibinfo{person}{Whitfield Diffie} {and}
  \bibinfo{person}{Martin Hellman}.} \bibinfo{year}{2006}\natexlab{}.
\newblock \showarticletitle{{New Directions in Cryptography}}.
\newblock \bibinfo{journal}{\emph{{IEEE Transactions on Information Theory}}}
  \bibinfo{volume}{22}, \bibinfo{number}{6} (\bibinfo{year}{2006}),
  \bibinfo{pages}{644--654}.
\newblock


\bibitem[\protect\citeauthoryear{Dobraunig, Eichlseder, Mendel, and
  Schlaffer}{Dobraunig et~al\mbox{.}}{2016}]%
        {Dobraunig2016}
\bibfield{author}{\bibinfo{person}{Christoph Dobraunig}, \bibinfo{person}{Maria
  Eichlseder}, \bibinfo{person}{Florian Mendel}, {and} \bibinfo{person}{Martin
  Schlaffer}.} \bibinfo{year}{2016}\natexlab{}.
\newblock \bibinfo{title}{{Ascon v1.2: Submission to the CAESAR Competition}}.
\newblock
  \bibinfo{howpublished}{\url{https://competitions.cr.yp.to/round3/asconv12.pdf}}.
\newblock
\newblock
\shownote{Date accessed: July 28, 2023.}


\bibitem[\protect\citeauthoryear{Fleischmann, Forler, and Lucks}{Fleischmann
  et~al\mbox{.}}{2012}]%
        {Fleischmann2012}
\bibfield{author}{\bibinfo{person}{Ewan Fleischmann},
  \bibinfo{person}{Christian Forler}, {and} \bibinfo{person}{Stefan Lucks}.}
  \bibinfo{year}{2012}\natexlab{}.
\newblock \showarticletitle{McOE: A Family of Almost Foolproof On-Line
  Authenticated Encryption Schemes}. In \bibinfo{booktitle}{\emph{Workshop on
  Fast Software Encryption}}. \bibinfo{pages}{196--215}.
\newblock


\bibitem[\protect\citeauthoryear{Guo, Iwata, Khairallah, Minematsu, and
  Peyrin}{Guo et~al\mbox{.}}{2018}]%
        {GuoNist}
\bibfield{author}{\bibinfo{person}{Chun Guo}, \bibinfo{person}{Tetsu Iwata},
  \bibinfo{person}{Mustafa Khairallah}, \bibinfo{person}{Kazuhiko Minematsu},
  {and} \bibinfo{person}{Thomas Peyrin}.} \bibinfo{year}{2018}\natexlab{}.
\newblock \bibinfo{title}{{Romulus v1.3: Submission to the NIST lightweight
  Competition}}.
\newblock
  \bibinfo{howpublished}{\url{https://romulusae.github.io/romulus/docs/Romulusv1.3.pdf}}.
\newblock
\newblock
\shownote{Date accessed: July 28, 2023.}


\bibitem[\protect\citeauthoryear{Jean, Nikoli{\'{c}}, Peyrin, and Seurin}{Jean
  et~al\mbox{.}}{2016}]%
        {Jean2016}
\bibfield{author}{\bibinfo{person}{J{\'e}r{\'e}my Jean}, \bibinfo{person}{Ivica
  Nikoli{\'{c}}}, \bibinfo{person}{Thomas Peyrin}, {and}
  \bibinfo{person}{Yannick Seurin}.} \bibinfo{year}{2016}\natexlab{}.
\newblock \bibinfo{title}{{Deoxys v1.41: Submission to the CAESAR
  Competition}}.
\newblock
  \bibinfo{howpublished}{\url{https://competitions.cr.yp.to/round3/deoxysv141.pdf}}.
\newblock
\newblock
\shownote{Date accessed: July 28, 2023.}


\bibitem[\protect\citeauthoryear{{Jean, J{\'e}r{\'e}my and Nikoli{\'{c}}, Ivica
  and Peyrin, Thomas}}{{Jean, J{\'e}r{\'e}my and Nikoli{\'{c}}, Ivica and
  Peyrin, Thomas}}{2014}]%
        {Jean2014}
\bibfield{author}{\bibinfo{person}{{Jean, J{\'e}r{\'e}my and Nikoli{\'{c}},
  Ivica and Peyrin, Thomas}}.} \bibinfo{year}{2014}\natexlab{}.
\newblock \showarticletitle{{Tweaks and Keys for Block Ciphers: The TWEAKEY
  Framework}}. In \bibinfo{booktitle}{\emph{{Proceedings of the 34th Annual
  International Cryptology Conference on Advances in Cryptology}}}.
  \bibinfo{pages}{274--288}.
\newblock


\bibitem[\protect\citeauthoryear{Kaliski}{Kaliski}{1998}]%
        {Kaliski1998}
\bibfield{author}{\bibinfo{person}{B. Kaliski}.}
  \bibinfo{year}{1998}\natexlab{}.
\newblock \bibinfo{title}{{PKCS {\#}7: Cryptographic Message Syntax Version
  1.5}}.
\newblock
  \bibinfo{howpublished}{\url{https://datatracker.ietf.org/doc/html/rfc2315}}.
\newblock
\newblock
\shownote{Date accessed: July 28, 2023.}


\bibitem[\protect\citeauthoryear{Krawczyk}{Krawczyk}{2001}]%
        {Krawczyk2001}
\bibfield{author}{\bibinfo{person}{Hugo Krawczyk}.}
  \bibinfo{year}{2001}\natexlab{}.
\newblock \showarticletitle{{The Order of Encryption and Authentication for
  Protecting Communications (or: How Secure Is SSL?)}}. In
  \bibinfo{booktitle}{\emph{{Proceedings of the 38th Annual International
  Cryptology Conference on Advances in Cryptology}}}.
  \bibinfo{pages}{310--331}.
\newblock


\bibitem[\protect\citeauthoryear{Liskov, Rivest, and Wagner}{Liskov
  et~al\mbox{.}}{2002}]%
        {Liskov2002}
\bibfield{author}{\bibinfo{person}{Moses Liskov}, \bibinfo{person}{Ronald~L.
  Rivest}, {and} \bibinfo{person}{David Wagner}.}
  \bibinfo{year}{2002}\natexlab{}.
\newblock \showarticletitle{Tweakable Block Ciphers}. In
  \bibinfo{booktitle}{\emph{{Proceedings of the 22nd Annual International
  Cryptology Conference on Advances in Cryptology}}}. \bibinfo{pages}{31--46}.
\newblock


\bibitem[\protect\citeauthoryear{Peyrin and Seurin}{Peyrin and Seurin}{2016}]%
        {Peyrin2016}
\bibfield{author}{\bibinfo{person}{Thomas Peyrin} {and}
  \bibinfo{person}{Yannick Seurin}.} \bibinfo{year}{2016}\natexlab{}.
\newblock \showarticletitle{Counter-in-Tweak: Authenticated Encryption Modes
  for Tweakable Block Ciphers}. In \bibinfo{booktitle}{\emph{Proceedings of the
  36th Annual International Cryptology Conference on Advances in Cryptology}}.
  \bibinfo{pages}{33--63}.
\newblock


\bibitem[\protect\citeauthoryear{Shoup}{Shoup}{1996}]%
        {Shoup1996}
\bibfield{author}{\bibinfo{person}{Victor Shoup}.}
  \bibinfo{year}{1996}\natexlab{}.
\newblock \showarticletitle{On Fast and Provably Secure Message Authentication
  Based on Universal Hashing}. In \bibinfo{booktitle}{\emph{{Proceedings of the
  16th Annual International Cryptology Conference on Advances in Cryptology}}}.
  \bibinfo{pages}{313--328}.
\newblock


\end{thebibliography}

\end{document}